\documentclass[aps,prl,twocolumn,preprintnumbers,amsmath,amssymb,superscriptaddress]{revtex4-1}
\usepackage{graphicx}
\usepackage{subfigure}
\usepackage{epsfig}
\usepackage{dcolumn}
\usepackage{bm}
\usepackage{ulem}
\usepackage{color}
\usepackage{multirow}
\usepackage{slashed}

\def\be{\begin{equation}}       \def\ee{\end{equation}}
\def\bea{\begin{eqnarray}}      \def\eea{\end{eqnarray}}
\def\ba{\begin{array}}
\def\ea{\end{array}}
\def\bnum{\begin{enumerate} }
\def\enum{\end{enumerate}}

\def\nn{\nonumber}

\def\=>{\Rightarrow}
\def\>{\rightarrow}

\def\eye2{Fathbb{I}}

\def\Eq#1{Eq.~(\ref{#1})}
\def\Fig#1{Fig.~\ref{#1}}

\newcommand{\no}{\nonumber}

\renewcommand{\>}{\rangle}

\begin{document}

\title{Emergence of supersymmetric quantum electrodynamics}

\author{Shao-Kai Jian}
\affiliation{Institute for Advanced Study, Tsinghua University, Beijing 100084, China}

\author{Chien-Hung Lin}
\affiliation{Department of Physics, University of Alberta, Edmonton, Alberta T6G 2E1, Canada}

\author{Joseph Maciejko}
\affiliation{Department of Physics, University of Alberta, Edmonton, Alberta T6G 2E1, Canada}
\affiliation{Theoretical Physics Institute, University of Alberta, Edmonton, Alberta T6G 2E1, Canada}
\affiliation{Canadian Institute for Advanced Research, Toronto, Ontario M5G 1Z8, Canada}

\author{Hong Yao}
\affiliation{Institute for Advanced Study, Tsinghua University, Beijing 100084, China}

\begin{abstract}
Supersymmetric (SUSY) gauge theories such as the Minimal Supersymmetric Standard Model play a fundamental role in modern particle physics, but have not been verified so far in nature. Here, we show that a SUSY gauge theory with dynamical gauge bosons and fermionic gauginos emerges naturally at the pair-density-wave (PDW) quantum phase transition on the surface of a correlated topological insulator (TI) hosting three Dirac cones, such as the topological Kondo insulator SmB$_6$. At the quantum tricritical point between the surface Dirac semimetal and nematic PDW phases, three massless bosonic Cooper pair fields emerge as the superpartners of three massless surface Dirac fermions. The resulting low-energy effective theory is the supersymmetric XYZ model, which is dual by mirror symmetry to $\mathcal{N}$=2 supersymmetric quantum electrodynamics (SQED) in 2+1 dimensions, providing a first example of emergent supersymmetric gauge theory in condensed matter systems. Supersymmetry allows us to determine exactly certain critical exponents and the optical conductivity of the surface states at the strongly coupled tricritical point, which may be measured in future experiments.
\end{abstract}
\date{\today}
\maketitle

Spacetime supersymmetry was introduced more than forty years ago as a means to resolve fundamental issues in particle physics such as the hierarchy problem~\cite{wessbook, gervais1971,wess1974, dimopoulos1981}, but has not been discovered yet. Amazingly, many beautiful theories originating in high-energy physics may be realized and tested in condensed matter systems; for instance, 3D Weyl fermions~\cite{ashvin2011,xu2011,burkov2011} were discovered recently in solid state materials~\cite{weng2015,hasan2015,xu2015,lv2015,yang2015}. One may wonder whether SUSY can be realized in quantum materials. Indeed, it was proposed that SUSY can emerge at quantum criticality in Bose-Fermi lattice models~\cite{lee2007,yu2010} and at the boundary of topological materials~\cite{grover2014,ponte2014,zerf2016,li2016}, as well as at multicritical points in low-dimensional systems~\cite{friedan1984,foda1988, huijse2015}. It was further shown in Ref. \cite{jian2015} that SUSY in 3+1D can emerge at superconducting quantum critical points in ideal Weyl semimetals~\cite{ruan2016a,ruan2016b}.

However, all known examples only realize the simplest type of emergent SUSY: the Wess-Zumino theory~\cite{wess1974}, which contains a single SUSY multiplet of matter fields (one scalar and one fermion). It is highly desirable to know whether richer types of SUSY can emerge in condensed matter systems, such as theories with dynamical gauge fields. Here, we theoretically show that the nematic PDW tricritical point on the surface of a correlated TI \cite{kane2010,qi2011} with three Dirac cones can realize a SUSY gauge theory. At this tricritical point, three Dirac fermions and three complex bosons form mutual superpartners and are described by the so-called XYZ model~\cite{strassler2003}, which is dual by mirror symmetry to $\mathcal{N}$=2 SQED~\cite{aharony1997,deboer1997,deboer1997b}.
\begin{figure}[b]
\subfigure{\includegraphics[height=3.9cm]{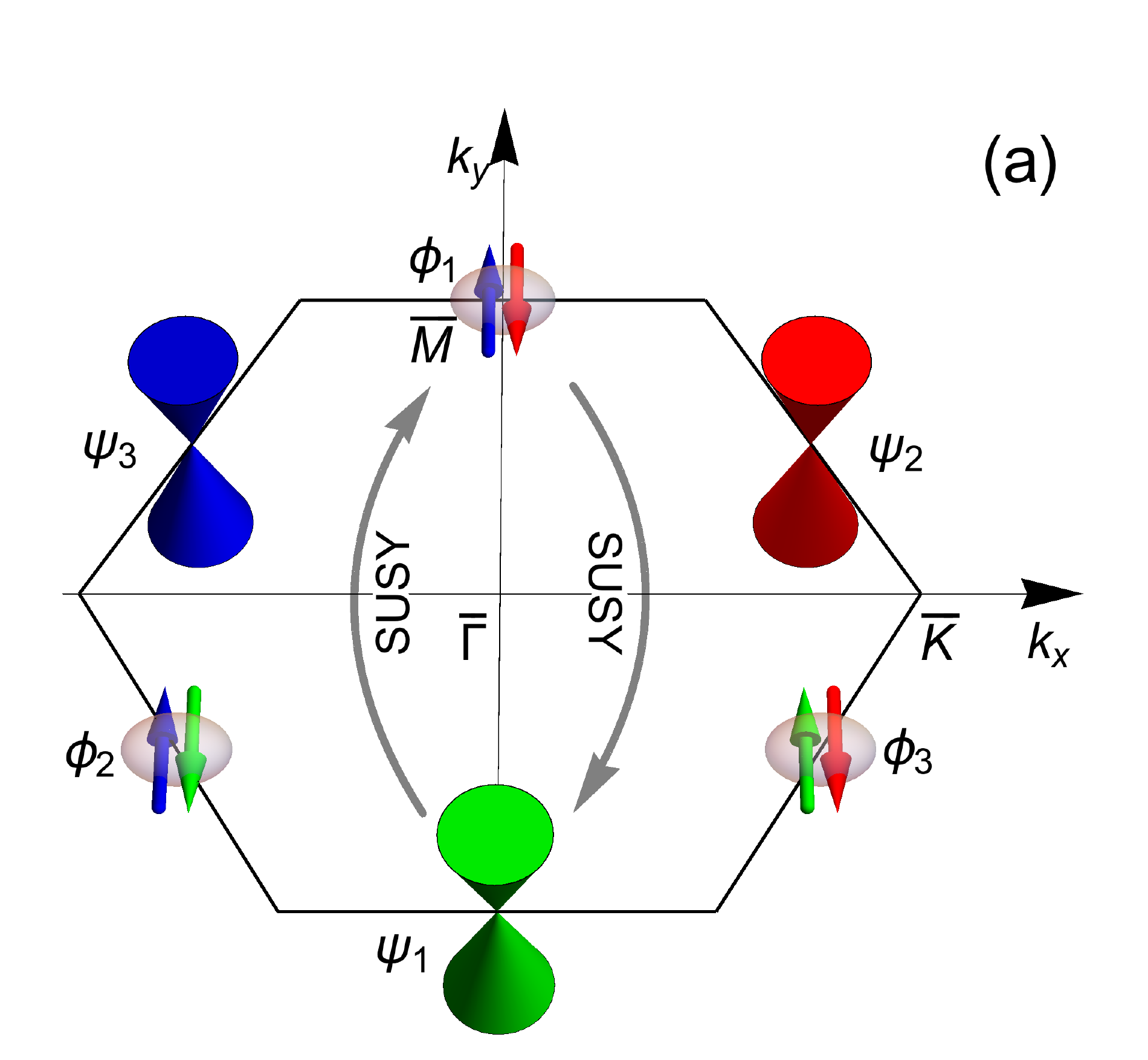}\label{BZ}}~~
\subfigure{\includegraphics[height=3.8cm]{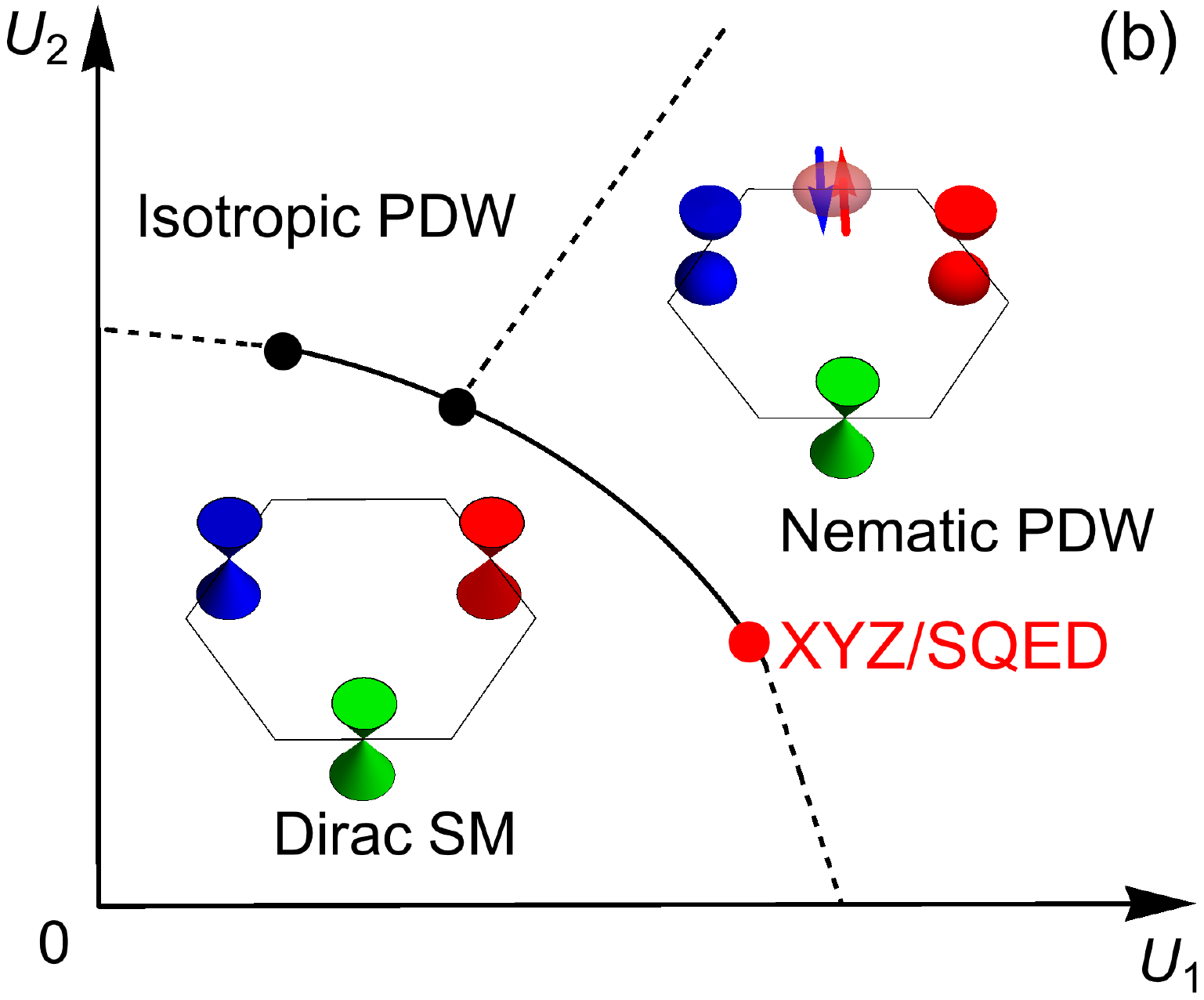}\label{phase}}
\vspace{-0.4cm}
\caption{(a) Hexagonal surface Brillouin zone of a TI with three Dirac cones $\psi_1$, $\psi_2$, and $\psi_3$, like in SmB$_6$. The three intervalley PDW order parameters are labeled by $\phi_1$, $\phi_2$, and $\phi_3$, which can be rotated into Dirac fermions by a SUSY transformation at the nematic PDW tricritical point. (b) Schematic quantum phase diagram. $U_1,U_2$ represent combinations of the couplings $r,u,u',u''$ in the Landau-Ginzburg theory (\ref{LGtheory}). Solid (dashed) lines represent second- (first-) order transitions. The red circle represents the tricritical point between the Dirac SM and nematic PDW phases, where SUSY of the XYZ/SQED type emerges. }
\end{figure}

An ideal candidate correlated TI to possibly realize this new type of SUSY is SmB$_6$, which is proposed to be a topological Kondo insulator~\cite{dzero2010,dzero2016} with three degenerate Dirac cones on its (111) surface protected by time-reversal and crystal symmetries~\cite{dai2013,ye2013}. (Another candidate with similar properties is YbB$_6$~\cite{weng2014}.) Experiments on surface electronic structure~\cite{xu2013,jiang2013,neupane2013}, transport properties~\cite{wolgast2013,shiyan2016}, and quantum oscillations~\cite{li2014,suchitra2015} in SmB$_6$ all indicate conducting surface Dirac cones but an insulating bulk. To realize three complex bosons as superpartners of the three surface Dirac fermions, we consider the surface quantum phase transition into a PDW phase with three complex order parameters.

From the effective theory describing the PDW transition on the TI surface hosting three Dirac cones, we analyze the possible phases of the model at the mean-field level and find two PDW phases distinguished by the lattice rotational symmetry. In the nematic PDW phase, which breaks rotational symmetry spontaneously, there is a tricritical point separating first- and second-order PDW phase transitions. A renormalization group (RG) analysis reveals that the SUSY XYZ model, and thus $\mathcal{N}=2$ SQED, emerges at this tricritical point. To the best of our knowledge, this is the {\it first} example of emergent SQED in quantum materials. We calculate certain critical exponents and the optical conductivity of the surface states exactly, which may be tested in future experiments on correlated TIs with PDW transitions.

{\bf Effective field theory:}
The hexagonal surface Brillouin zone (BZ) of a TI with C$_3$ symmetry (e.g., Bi$_2$Se$_3$ and SmB$_6$) contains four time-reversal invariant (TRI) points: the $\bar{\Gamma}$ point, and three $\bar{M}$ points related by symmetry [Fig.~\ref{BZ}]. There are two different types of surface states: a single Dirac cone at the $\bar{\Gamma}$ point, like on the (001) surface of Bi$_2$Te$_3$ and Bi$_2$Se$_3$~\cite{kane2010,qi2011}, or three Dirac cones at the $\bar{M}$ points, like on the (111) surface of SmB$_6$~\cite{dai2013,ye2013} and YbB$_6$~\cite{weng2014}. Here we consider the latter case. We further impose the reflection symmetry $\mathcal{M}_x$ ($x\!\rightarrow\! -x$) which is respected in SmB$_6$. As a result, the TI surface has C$_{3v}$ symmetry. The three Dirac fermions located at these TRI points are denoted by $\psi_{1,2,3}$ [Fig.~\ref{BZ}]. The little group for $\psi_1$ is generated by time reversal $\mathcal{T}$ and reflection $\mathcal{M}_x$. The low-energy theory of the surface Dirac semimetal (DSM) is dictated by symmetry and given by
\bea
	\mathcal{L}_f=\sum_{i=1}^3\psi^\dag_i (\partial_\tau+h^f_i)\psi_i,
\eea
where $\tau$ is imaginary time and $h^f_1= -i \sigma^y v_x \partial_x +i \sigma^x v_y \partial_y $ is a Dirac-like Hamiltonian with $v_i$ the fermion velocities and $\sigma^i$ the Pauli spin matrices; $h^f_2$ and $h^f_3$ are obtained from $h^f_1$ by rotations~\cite{supp}. Though there is no symmetry to enforce $v_x\!=\!v_y$, velocity isotropy emerges at the PDW transitions discussed below. We assume that the chemical potential is exactly at the Dirac points, namely at stoichiometry. 
By contrast with the (111) surface considered here, on the (001) surface of SmB$_6$ the three Dirac points are not at equal energy \cite{say2015}. 

We consider the system near PDW criticality, where pairing is between different cones and the PDW order parameters are, e.g., $\phi_1\propto \psi_2 \sigma^y \psi_3$ [Fig.~\ref{BZ}]. PDW ordering possesses finite momentum but does not spontaneously break time-reversal symmetry~\cite{FF,LO,berg2007,berg2009,ashvin2009}. The quantum Landau-Ginzburg Lagrangian for the PDW order parameters is constrained by symmetry and reads~\cite{supp}
\bea
	\mathcal{L}_b &=& \sum_{i=1}^3 \phi_i^\ast (-\partial^2_\tau+ h_i^b) \phi_i +V_b, \\
	V_b &=& r\sum_{i=1}^3 |\phi_i|^2 + u (|\phi_1|^2 |\phi_2|^2 + \text{c.p.}) \nn\\
	&& + u' \big[(\phi_1^{\ast2} \phi_2^2+ \text{h.c.} )+ \text{c.p.}\big] + u''\sum_{i=1}^3 |\phi_i|^4,\label{LGtheory}
\eea
where $r,u,u',u''$ are phenomenological constants, c.p. denotes cyclic permutations, and $h_2^b$, $h_3^b$ can be obtained from $h_1^b=-c_x^2 \partial_x^2 - c_y^2 \partial_y^2$ by rotations. Terms linear in spatial derivatives are forbidden due to time-reversal symmetry (we also implicitly assumed particle-hole symmetry to rule out terms linear in time derivative), and higher order terms omitted in $\mathcal{L}_b$ are irrelevant in the RG sense. Boson and fermion velocities are initially different, but flow to a common value in the infrared as discussed later in the text. We hereafter assume $u'<0$ since the Josephson coupling between different condensates normally minimizes their superconducting phase difference. Moreover, the Dirac fermions and PDW order parameter fluctuations are coupled:
\bea
	\mathcal{L}_{bf}= g(\phi_1 \psi_2 \sigma^y \psi_3+ \text{c.p.})+\text{h.c.}, \label{bf}
\eea
where $g$ is a coupling constant.

{\bf Mean-field analysis:}
To facilitate the analysis of the possible PDW phases at the mean-field level, we rewrite the boson potential as
\bea
V_b=u^{\prime\prime}\left(\sum_{i=1}^3|\phi_i|^2\right)^2\!+\! (u\!-\!2u'') (|\phi_1|^2 |\phi_2|^2\!+\!\text{c.p.}), \label{potential}
\eea
where we have implicitly absorbed the $u'$ term into the $u$ term (i.e., $u\!+\!2u' \!\to\! u$) because the phase differences between different condensates (Leggett modes) are gapped in the ordered phase~\cite{leggett1966}. For now, we neglect the Dirac fermions in the lowest order approximation.  The sign of the second (anisotropic) term in \Eq{potential} is crucial to determine which PDW ground state is preferred. In the PDW ordered phases, the mass term is negative $r\!<\!0$.  When $u\!-\!2u''\!<\!0$ and $u+u''>0$, the anisotropic term in the potential favors the ordering $|\phi_i|\!=\!\sqrt{\frac{|r|}{2(u+u'')}}$ and $\phi_1\!=\!\phi_2\!=\!\phi_3$, which we denote the isotropic PDW (IPDW) phase because it preserves the crystalline C$_{3v}$ symmetry. In the IPDW phase, all three surface Dirac fermions are gapped by pairing. On the other hand, $u\!-\!2u''\!>\!0$ and $u''>0$ favors a qualitatively different type of PDW ordering: only one component condenses with $|\phi_i|\!=\!\sqrt{\frac{|r|}{2u''}}$ while the other two $\phi_{j\neq i}$ vanish. We call this phase the nematic PDW (NPDW) because it breaks C$_{3v}$ spontaneously. There is no secondary charge-density-wave order formed in the NPDW phase. In the NPDW phase, only two Dirac points are gapped and one remains massless. For the special case $u\!=\!2u''$, the theory describes a bicritical point where the DSM-IPDW and DSM-NPDW phase boundaries meet [Fig.~\ref{phase}].

In the analysis above, we have implicitly assumed that the transition between the DSM and PDW phases is continuous. However, it is always possible for a transition to be discontinuous. We thus also consider the possibility of first-order PDW transitions as well as tricritical points between the first- and second-order transitions. The transition into the nematic PDW phase (i.e., $u\!-\!2u''\!>\!0$) should be first-order when $u''\!<\!0$, in which case a sixth-order term like $w(\sum_{i=1}^3|\phi_i|^2)^3$ with $w\!>\!0$ should be added to $V_b$ to stabilize the free energy. $u''\!=\!0$ is thus a tricritical point between the continuous ($u''\!>\!0$) and first-order ($u''\!<\!0$) transitions into the nematic PDW phase [\Fig{phase}]. Similarly, $u+u''=0$ describes a tricritical point between the continuous transition ($u+u''\!>\!0$) and the first-order transition ($u''+u\!<\!0$) into the isotropic PDW phase.

We have identified three multicritical points through the mean-field analysis above: one bicritical, and two tricritical. In the remainder of the paper we analyze the emergent low-energy, long-wavelength properties at these multicritical points. Remarkably, the tricritical point into the nematic PDW phase features an emergent SUSY of the XYZ/SQED type, as discussed below.

{\bf Effective theory of the bicritical point: }
We first explore universal properties of the continuous DSM-PDW transition (i.e., $r\!=\!0$) via a one-loop RG analysis in $4\!-\!\epsilon$ spacetime dimensions (the physical dimension corresponds to $\epsilon=1$)~\cite{supp}. At this transition, we find that the anisotropy in fermion and boson velocities vanishes, i.e., $c_x\!=\!c_y\!\equiv \!c$ and $v_x\!=\!v_y\!\equiv\!v$ at low energies and long distances. Moreover, they flow to a common value $c\!=\!v$ in the infrared such that Lorentz symmetry emerges at the continuous PDW transition, no matter whether the PDW phase is isotropic or nematic. Even though emergent Lorentz symmetry was previously observed at various quantum critical points involving Dirac fermions~\cite{jian2015, vafek2002, isobe2012, roy2015, FIQCP2015}, it is more exotic here because it involves an odd number of Dirac cones on the surface of a correlated TI. This emergent Lorentz symmetry allows us to set $c\!=\!v\!=\!1$ in discussing the PDW quantum critical points.

From the RG equations for the coupling constants $g,u,u',u''$,
\bea
	\frac{d g^2}{dl} &=& \epsilon g^2-\frac{3\pi}{2} g^4, \no\\
	\frac{du}{dl} &=&\epsilon u -\pi g^2 u+ \pi g^4- \frac{\pi}{2} (3u^2+16u'^2+8uu''),\no \\
	\frac{du'}{dl} &=&\epsilon u'  -\pi g^2 u'- \pi(u'^2+2uu'+2u'u''), \no\\
	\frac{du''}{dl} &=& \epsilon u'' -\pi g^2 u'' + \frac{\pi}{2} g^4 - \frac{\pi}{2}(u^2+4u'^2+10u''^2),\no
\eea
we find a unique stable fixed point at $g^{2}_\text{st}\!=\!\frac{2}{3\pi}\epsilon$, $u^{\prime}_\text{st}\!=\!0$, and $u_\text{st}\!=\!2u^{\prime\prime}_\text{st}\!=\!\frac{1+\sqrt{57}}{21\pi}\epsilon$, where the subscript ``$\text{st}$'' means ``stable". At this stable fixed point, the boson potential becomes $V_b=u^{\prime\prime}_\text{st}(\sum_{i=1}^3|\phi_i|^2)^2$ and has an emergent $\mathrm{SO(6)}$ symmetry. However, the full theory only has the reduced $\mathrm{U(1)}\times\mathrm{C}_{3v}$ symmetry due to the finite fermion-boson coupling $g_\text{st}$.

Based on our earlier analysis, the fixed point with $u_\text{st}\!-\!2u''_\text{st}\!=\!0$ corresponds to a bicritical point where three phases (DSM, IPDW, and NPDW) meet. However, this multicritical point is a novel one as it has only one relevant direction (the mass term $r$). The term proportional to $u\!-\!2u''$ in Eq. (\ref{potential}) is dangerously irrelevant and ground states on the ordered side crucially depend on its sign.

{\bf Emergent XYZ/SQED at the NPDW tricritical point:} Besides the stable fixed point discussed above, the RG equations also support another (unstable) fixed point at $g^{2}_{\text{susy}}\!=\!u_{\text{susy}}\!=\!\frac{2}{3\pi}\epsilon$ and $u^{\prime}_{\text{susy}}\!=\!u^{\prime\prime}_{\text{susy}}\!=\!0$. The fixed point action is invariant under the SUSY transformations $\delta \phi_i \!=\!\sqrt{2} \xi \psi_i$, $\delta \psi_1 \!=\! i\sqrt{2} \sigma^\mu \bar{\xi} \partial_\mu \phi_1 \!+\! g\sqrt{2} \xi \phi_2\phi_3$, and $\delta \psi_{2,3}$ are obtained by permutations of $\delta \psi_1$, where the infinitesimal transformation parameters $\xi, \bar{\xi}$ are Grassmann-valued two-component spinors, and $\sigma^0\! =\!-\!I$ with $I$ the identity matrix.

Remarkably, this fixed point is described by a new type of SUSY qualitatively different from all previously predicted in condensed matter. The bosonic PDW fields $\phi_i$ and Dirac fermions $\psi_i$ combine into three chiral superfields $\Phi_i \!=\! \phi_i \!+\! \sqrt{2} \theta^\alpha\sigma^y_{\alpha\beta} \psi^\beta_i \!+\!...$, $i=1,2,3$, where $\theta$ is a Grassmann-valued two-component spinor and $\alpha,\beta$ are (pesudo-)spin indices. Intravalley pairing would have resulted in three decoupled copies of the $\mathcal{N}=2$ Wess-Zumino theory with superpotential $\Phi_i^3$ studied previously~\cite{lee2007,yu2010,grover2014,ponte2014,zerf2016,jian2015}. By contrast, in the intervalley pairing scenario considered here the three valleys are strongly coupled via the superpotential $\Phi_1\Phi_2\Phi_3$~\cite{supp}, and the resulting theory is known as the XYZ model. It flows in the infrared to a strongly coupled fixed point, which is dual via mirror symmetry---a SUSY version of the Peskin-Dasgupta-Halperin or particle-vortex duality~\cite{peskin1978,dasgupta1981,fisher1989}---to the infrared fixed point of $\mathcal{N}=2$ SQED~\cite{aharony1997,deboer1997,deboer1997b}. The latter is a theory of a vector superfield $V$ and two chiral superfields $Q,\tilde{Q}$, playing the role of gauge field and matter field in the ``vortex'' theory, respectively. In addition to an emergent bosonic gauge field $A_\mu$, the vector superfield $V$ also contains a fermionic gaugino $\lambda$.

We now show that the XYZ/SQED fixed point with $u'\!=\!u''\!=\!0$ and $u\!>\!0$ corresponds to the tricritical point that separates the continuous and first-order transitions into the nematic PDW phase. Linearizing the RG equations for $g,u,u',u''$ near the SUSY fixed point, we can determine the eigenoperators at this fixed point and their eigenvalues, which are ($-\frac73$,$-$1,$-$1,1). The positive eigenvalue indicates that there is one relevant direction (besides the relevant direction of $r$). Consequently, the XYZ/SQED fixed point is unstable. This is consistent with our mean-field analysis of the tricritical point on the transition boundary between the DSM and NPDW phases, which is reached by tuning two parameters.

The emergence at the NPDW tricritical point of the XYZ SUSY, which is dual to SQED, may be intuitively understood as follows. Heuristically, for a fermionic quantum critical point with the same number of Dirac fermions and complex order parameters to be possibly supersymmetric, one necessary condition is that the coupling among different bosonic order parameters should avoid flowing to infinity (namely it should not be relevant), otherwise the number of remaining effective gapless bosonic modes would be less than the number of fermionic ones. The nematic PDW breaks the U(1) gauge symmetry, as well as the lattice $C_3$ symmetry which is effectively a U(1) symmetry at criticality due to the irrelevance of anisotropic terms. From a symmetry point of view, it is thus natural to expect two gapless complex bosonic modes at a generic NPDW quantum critical point. However, the low-energy theory has three gapless Dirac fermions. To have a chance of being supersymmetric, the quantum phase transition must be tuned to a multicritical point such that a third complex bosonic mode remains gapless. Here this multicritical point is the NPDW tricritical point.

Like the Wess-Zumino model, the XYZ model enjoys an $R$-symmetry~\cite{wessbook,aharony1997}. The $R$-charge of the superpotential $\Phi_1\Phi_2\Phi_3$ should be 2, i.e., $\sum_{i=1}^3 \mathcal{R}(\Phi_i) \!=\!2$, where $\mathcal{R}(\Phi_i)$ denotes the $R$-charge of the superfield $\Phi_i$. The assignment of $R$-charge for the superfield $\Phi_i$ is simple owing to the rotational symmetry in our case: they should be equal, $\mathcal{R}(\Phi_i)\!=\!\frac{2}{3}$. For a chiral superfield, scaling dimension is exactly equal to the $R$-charge~\cite{seiberg1994, aharony1997} in 2+1 dimensions. We thus obtain the exact scaling dimensions of the bosonic order parameter fluctuations and Dirac fermions as $\Delta_\phi\!=\! \frac{2}{3}$ and $\Delta_\psi\!=\!\Delta_\phi\!+\!\frac12=\frac76$, respectively. Setting $\epsilon\!=\!1$, our one-loop RG result $\Delta_\phi\!=\!\frac12\!+\!\frac{\epsilon}{6}=\frac23$ for the boson scaling dimension is consistent with the exact result. 
Accordingly, the order parameter anomalous dimension or critical exponent $\eta$ is $\frac13$. On the other hand, the correlation length exponent $\nu$ is related to the scaling dimension of nonchiral fields $|\phi_i|^2$, and cannot be simply related to the $R$-charge. We obtain $\nu \!=\! \frac12\!+\! \frac\epsilon4+\mathcal{O}(\epsilon^2)$ at the one-loop level~\cite{supp}.

{\bf Experimental signatures of SUSY at the NPDW tricritical point:} Owing to the strong constraints imposed by $\mathcal{N}=2$ superconformal symmetry at the XYZ/SQED fixed point, several dynamical properties can be obtained exactly~\cite{wwk2015,nishioka2013} despite the presence of strong interactions at this fixed point. According to linear response theory, the optical conductivity at frequency $\omega$ is given by the current-current correlation function,
\bea\label{kubo}
	\sigma(\omega)= \frac{e^2}{\hbar}\frac{1}{i\omega}\big\langle J_x(\omega) J_x(-\omega) \big\rangle,
\eea
where $\frac{e^2}{\hbar}$ is the quantum of conductance. The current-current correlation function is highly constrained by conformal symmetry through the conformal Ward identity~\cite{nishioka2013,osborn1994}.

We now compute the exact optical conductivity at the NPDW tricritical point. Utilizing the $R$-symmetry of $\mathcal{N}\!=\!2$ superconformal field theories in 2+1 dimensions, one can find~\cite{wwk2015} that $\sigma_0(\omega)=\frac54 \tau_{RR} \frac{e^2}{\hbar}$, where $\sigma_0(\omega)$ is the optical conductivity at zero temperature and $\tau_{RR}$ is the dimensionless coefficient of the two-point correlation function of the $R$-current~\cite{supp,closset2013,imamura2011,imamura2012}. Assuming $\omega\ll\Lambda$ where $\Lambda$ represents microscopic energy scales above which quantum critical behavior ceases to exist, the zero-temperature optical conductivity $\sigma_0(\omega)=\sigma_0$ is a universal constant independent of frequency that characterizes the universality class of the transition, just like critical exponents~\cite{damle1997}. In an $\mathcal{N}=2$ SUSY field theory with only chiral superfields, $\tau_{RR}$ is given by an integral that depends only on the $R$-charge of the chiral superfields~\cite{nishioka2013}. As mentioned before, owing to the C$_3$ rotational symmetry relating the three chiral superfields $\Phi_i$, we obtain $\mathcal{R}(\Phi_i)=\frac23$, $i=1,2,3$. This is the same as the $R$-charge of the chiral superfield in the Wess-Zumino model~\cite{aharony1997}. As a result, $\tau_{RR}$ at the XYZ/SQED fixed point is simply three times that in the Wess-Zumino model, which was evaluated analytically in Ref.~\cite{wwk2015}. The exact zero-temperature optical conductivity at the nematic PDW tricritical point is thus given by
\bea
\sigma_0(\omega)=\frac{15}{243}\left(16-\frac{9\sqrt{3}}{\pi}\right) \frac{e^2}{\hbar}\approx 0.681 \frac{e^2}{\hbar},
\eea
which may be tested in future experiments.

Other experimental signatures include various critical exponents already mentioned such as the fermion/boson anomalous dimension $\eta=\frac{1}{3}$, which is exact owing to SUSY, and the correlation length exponent $\nu\approx 0.75$. The exact value of $\eta$ implies that the local electronic density of states $\rho(\omega)$ scales as $|\omega|^{4/3}$ at low energies~\cite{ponte2014}, which can be measured by scanning tunneling microscopy (STM).

{\bf Concluding remarks:} We have shown that a novel type of SUSY emerges at the nematic PDW tricritical point on the surface of a correlated TI that hosts three Dirac cones, like SmB$_6$. At this tricritical point, the three surface Dirac fermions and three complex bosons (corresponding to PDW order parameter fluctuations) are described by the so-called XYZ model, which is dual in the low-energy and long-wavelength limit to a SUSY gauge theory, $\mathcal{N}=2$ SQED. As such, our result also provides a direct physical setting for the investigation of mirror symmetry in condensed matter systems~\cite{sachdev2010,hook2013,hook2014}. This is an area of increased recent activity~\cite{kachru2015,kachru2016} owing to its connection with a series of recently proposed dualities in 2+1 dimensions~\cite{son2015, wang2015, metlitski2016, wang2016, mross2016, karch2016, seiberg2016, murugan2016}, with applications to a wide range of problems in contemporary condensed matter physics such as quantum spin liquids, topological phases, and the half-filled Landau level.

We have also predicted various critical exponents and the zero-temperature optical conductivity at the nematic PDW tricritical point, which could be tested in future experiments. If the SUSY proposed in the present paper is realized in condensed matter systems, it would help to determine various quantities {\it nonperturbatively}, such as the critical exponent $\nu$ of SQED in 2+1 dimensions, which are theoretically known only perturbatively (or numerically by bootstrap calculations~\cite{bobev2015a,bobev2015b}). We hope the present results will stimulate the theoretical and experimental search for various types of emergent SUSY and, more generally, emergent phenomena in condensed matter systems~\cite{wenbook}.

{\it Acknowledgements:} We would like to thank K. Sun, G. Torroba, and W. Witczak-Krempa for helpful discussions. This work was in part supported by by the NSFC under Grant No. 11474175 at Tsinghua University (SKJ and HY), and by NSERC grant \#RGPIN-2014-4608, the CRC Program, CIFAR, and the University of Alberta (CHL and JM). JM would like to thank the hospitality of the Institute for Advanced Study, Tsinghua University where this work was initiated.

\begin{widetext}

\section{Supplemental Material}
\renewcommand{\theequation}{S\arabic{equation}}
\setcounter{equation}{0}
\renewcommand{\thefigure}{S\arabic{figure}}
\setcounter{figure}{0}
\renewcommand{\thetable}{S\arabic{table}}
\setcounter{table}{0}

\subsection{A. Effective field theory at criticality}

\begin{figure}[b]
	\includegraphics[width=7cm]{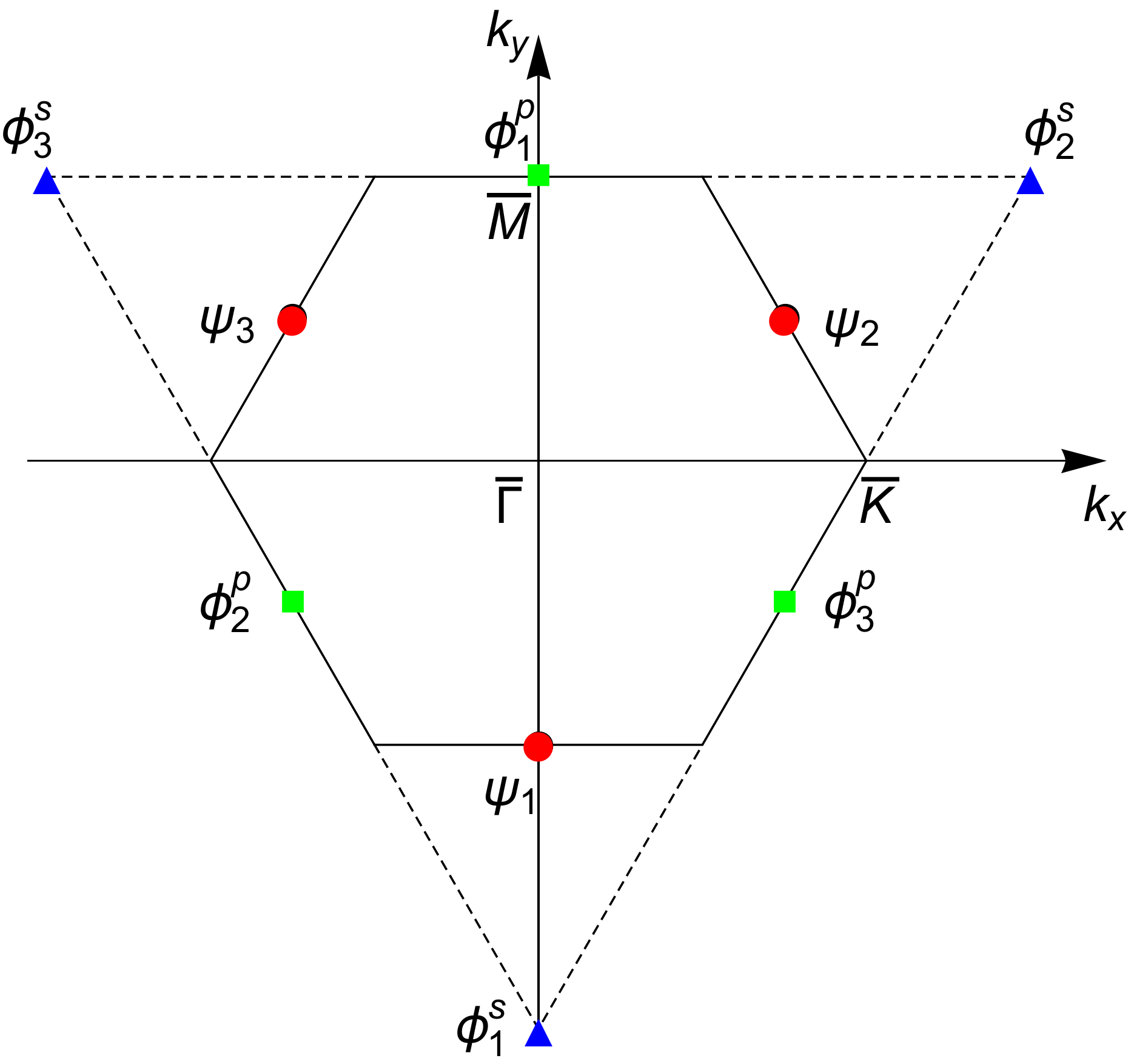}
	\caption{The red circles stand for surface Dirac cones at time-reversal invariant points. Three green rectangles denote the momenta for intervalley PDW pairing. Three blue triangles represent intravalley SC pairing. Note that SC pairing order has zero crystal momentum. \label{pairing}}
\end{figure}

The kinetic part of the Lagrangian for surface Dirac fermions is $\mathcal{L}_f = \sum_i \psi^\dag_i (-iw+h^f_i )\psi_i$, with $h_i^f$ given by
\bea
	h^f_1 &=& - v_y k_y \sigma^x+ v_x k_x \sigma^y,   \\
	h^f_2 &=& \left[\frac{\sqrt{3}}{4}(v_x- v_y) k_x - \left(\frac{3}{4}v_x + \frac{1}{4} v_y \right)k_y\right] \sigma^x+\left[\left(\frac{1}{4} v_x+ \frac{3}{4} v_y\right) k_x + \frac{\sqrt{3}}{4}(v_y-v_x)k_y\right] \sigma^y,  \\
	h^f_3 &=& \left[\frac{\sqrt{3}}{4}(v_y- v_x) k_x - \left(\frac{3}{4}v_x + \frac{1}{4} v_y \right)k_y\right] \sigma^x+\left[\left(\frac{1}{4} v_x+ \frac{3}{4} v_y\right) k_x + \frac{\sqrt{3}}{4}(v_x-v_y)k_y\right] \sigma^y,
\eea
where $i=1,2,3$ represent three valleys at the momentum points $(0,-1), (\frac{\sqrt{3}}{2},\frac{1}{2}), (-\frac{\sqrt{3}}{2},\frac{1}{2})$ in units of $\frac{2\sqrt{3}\pi}{3a}$ where $a$ is the lattice constant. $v_i$ are the two components of fermion velocity and $\sigma^i$ are the Pauli matrices. $h_1^f$ is dictated by the little group of $\psi_1$, and $h_2^f$, $h_3^f$ are obtained by the rotation operators $\hat R= e^{-i\frac{2\pi}{3} \hat L}\otimes e^{-i \frac{2\pi}{3}  \hat S} \otimes \hat \Lambda$, where $\hat L,\hat S, \hat \Lambda \!\equiv\!\left( \ba{cccc} 0 & 1 & 0 \\ 0 & 0 & 1 \\ 1 & 0 & 0 \ea \right) $ are generators in orbit, spin, and valley space, respectively.

The (pseudo-)spin-singlet pairing order parameter is $\phi^* \propto \psi (\sigma^y \otimes \lambda) \psi$, where $\psi$ denotes $(\psi_1,\psi_2,\psi_3)^T$, while $\lambda$ denotes valley pairing and satisfies $\lambda^T=\lambda$. Up to a phase, there are two possibilities for $\lambda$: intervalley pair-density-wave (PDW) pairing,
\bea
	\lambda^p_1 =  \left( \ba{cccc} 0 & 0 & 0 \\ 0 & 0 & 1 \\ 0 & 1 & 0 \ea \right),
	\lambda^p_2 =  \left( \ba{cccc} 0 & 0 & 1 \\ 0 & 0 & 0 \\ 1 & 0 & 0 \ea \right),
	\lambda^p_3 =  \left( \ba{cccc} 0 & 1 & 0 \\ 1 & 0 & 0 \\ 0 & 0 & 0 \ea \right),
\eea
and intravalley superconducting (SC) pairing,
\bea
	\lambda^s_1 =  \left( \ba{cccc} 1 & 0 & 0 \\ 0 & 0 & 0 \\ 0 & 0 & 0 \ea \right) ,
	\lambda^s_2 =  \left( \ba{cccc} 0 & 0 & 0 \\ 0 & 1 & 0 \\ 0 & 0 & 0 \ea \right),
	\lambda^s_3 =  \left( \ba{cccc} 0 & 0 & 0 \\ 0 & 0 & 0 \\ 0 & 0 & 1 \ea \right).
\eea
These matrices form two distinct three-dimensional reducible representations of the three-fold rotational symmetry.

The coupling between Dirac fermions and bosonic order parameters is given by
\bea\label{EqS6}
	\mathcal{L}_{fb}= \sum_{a,i} g_a (\phi_i^a \psi \sigma^y \lambda^a_i \psi+ \text{h.c.}),
\eea
where $\phi^a$ is the order parameter field and $a=p,s$ denotes intervalley PDW pairing or intravalley SC pairing. Below, we focus on intervalley PDW criticality and omit the superscript $p$ for simplicity.

The symmetry that constrains the theory is $\mathrm{C}_{3v}\times\mathrm{U(1)}\times\mathcal{P}\times\mathcal{T}$, where $\mathcal{P}$ denotes translation symmetry and $\mathcal{T}$ corresponds to time reversal symmetry. Each PDW order parameter possesses finite momentum of magnitude $\frac{2\sqrt{3}\pi}{3a}$ (Fig.~\ref{pairing}). As a result, translationally invariant quadratic terms must have the form $\phi_i^2$ or $|\phi_i|^2$. These terms are further decomposed into irreducible representations of $\mathrm{C}_{3v}\times\mathrm{U(1)}$, as shown in Table~\ref{representation}.
\begin{table}[t]
\caption{Irreducible representations of $\mathrm{C}_{3v}\times\mathrm{U(1)}$ formed by the PDW order parameters, where $w=e^{i 2\pi/3}$.\label{representation}}
\begin{tabular}{ |c|c|c| }
\hline
Representation & U(1) singlet 			& U(1) doublet \\
\hline
A$_1$               & $\sum_i |\phi_i|^2$ 	& $\sum_i \phi_i^2$ \\
\hline
\multirow{2}{1em}{E}  & $|\phi_1|^2+ w|\phi_2|^2+ w^2 |\phi_3|^2$  & $\phi_1^2+ w\phi_2^2+ w^2 \phi_3^2$    \\
 						 & $|\phi_1|^2+ w^2 |\phi_2|^2+ w |\phi_3|^2$  & $\phi_1^2+ w^2 \phi_2^2+ w \phi_3^2$\\
\hline
\end{tabular}
\end{table}
From Table~\ref{representation}, a symmetry-allowed potential for the PDW field must have the form
\bea
	V_b= r\sum_i |\phi_i|^2 + u (|\phi_1|^2 |\phi_2|^2 + \text{c.p.})+ u' [(\phi_1^{\ast2} \phi_2^2+ \text{h.c.} )+ \text{c.p.}] + u''\sum_i |\phi_i|^4,
\eea
where $r,u,u',u''$ are constants, c.p. denotes cyclic permutations and h.c. indicates Hermitian conjugation. The kinetic part of the boson Lagrangian is
$K_b= \sum_i \phi_i^\ast (w^2+ h_i^b) \phi_i$, with $h^b_i$ given by
\bea
	h^b_1 &=& c_x^2 k_x^2 + c_y^2 k_y^2, \\
	h^b_2 &=& c_x^2 \left(-\frac{1}{2} k_x+ \frac{\sqrt{3}}{2} k_y\right)^2 + c_y^2 \left(-\frac{\sqrt{3}}{2} k_x- \frac{1}{2} k_y\right)^2, \\
	h^b_3 &=& c_x^2 \left(-\frac{1}{2} k_x- \frac{\sqrt{3}}{2} k_y\right)^2 + c_y^2 \left(\frac{\sqrt{3}}{2} k_x- \frac{1}{2} k_y\right)^2,
\eea
where $c_x, c_y$ are boson velocities, and $h^b_2$ and $h^b_3$ are obtained from $h^b_1$ by rotation.

\subsection{B. Renormalization group analysis}

Here, we perform a one-loop renormalization group (RG) analysis at PDW criticality. Fast modes are integrated out to generate the flow equations. The interaction term in the action is
\bea
	S_{\text{int}}= \int d^d x\left[ \sum_{i} g (\phi_i \psi \sigma^y \lambda^p_i \psi+ \text{h.c.}) + u (|\phi_1|^2 |\phi_2|^2 + \text{c.p.})+ u' [(\phi_1^{\ast2} \phi_2^2+ \text{h.c.} )+ \text{c.p.}]+ u'' \sum_i |\phi_i|^4\right].
\eea
We assume that the velocity anisotropies $\delta c \!\equiv\! c_y\!-\! c_x$, $\delta v \!\equiv\! v_y\!-\!v_x$ are small compared to $v_i$ or $c_i$, i.e., $|\delta c|, |\delta v| \ll c_i, v_i$,
and perform calculations to leading order in the anisotropy. Feynman propagators for fermions and bosons are denoted by $S_i(p)$ and $D_i(p)$, respectively. The fermion $\Sigma_i(p)$ and boson $\Pi_i(p)$ self-energy renormalizations are given by
\bea
	\Sigma_1(p) &=& -\frac{1}{2} \times 2 \times g^2 \left[ \int_k \sigma^y [- S^T_3(k)] \sigma^y D_2(k-p) + \sigma^y [- S^T_2(k)] \sigma^y D_3(k-p) \right] \nn\\
	&=& \frac{g^2 \Lambda^{d-4}l}{(2\pi)^d} \Big[ \left(\frac{8\pi^2}{c_x (c_x+v_x)^2}- \frac{16\pi^2}{3c_x(c_x+v_x)^3} \delta v- \frac{8\pi^2(3c_x+v_x)}{3c_x^2(c_x+v_x)^3} \delta c \right) (-iw) \nn\\
	&& + \left( -\frac{8\pi^2(2c_x+v_x)}{3c_x(c_x+v_x)^2}+ \frac{2\pi^2(2c_x^2+21c_x v_x +7 v_x^2)}{15c_x v_x(c_x+v_x)^3} \delta v + \frac{8\pi^2(7c_x^2+6c_x v_x+2 v_x^2 )}{15c_x^2(c_x+v_x)^3}  \delta c \right) p_y \sigma^x \nn\\
	&& + \left( \frac{8\pi^2(2c_x+v_x)}{3c_x(c_x+v_x)^2}+\frac{2\pi^2(2c_x^2-3c_x v_x -v_x^2)}{3c_x v_x(c_x+v_x)^3} \delta v -\frac{8\pi^2}{3(c_x+v_x)^3} \delta c \right) p_x \sigma^y \Big], \\
	\Pi_1(p) &=& -\frac{1}{2} \times 2 \times (-1) \times g^2 \left[ \int_k \text{Tr}[(\sigma^y)^T S_2^T(k) \sigma^y S_3(k+p)]  \right] \nn\\
	&=& \frac{g^2 \Lambda^{d-4}l}{(2\pi)^d} \Big[ \left(\frac{2\pi^2}{v_x^3}- \frac{2\pi^2}{v_x^4} \delta v\right) w_p^2 + \left(\frac{2\pi^2}{v_x}+ \frac{\pi^2}{v_x^2} \delta v\right) p_x^2 + \left(\frac{2\pi^2}{v_x}- \frac{\pi^2}{v_x^2} \delta v\right) p_y^2 \Big],
\eea
where $\int_k\!\equiv\! \int \frac{d^d k}{(2\pi)^d}$, $\text{Tr}$ stands for the trace in (pseudo-)spin space, $\Lambda$ is an ultraviolet momentum cutoff, and $l\!>\!0$ is the flow parameter. The self-energies for the other two valleys $\Sigma_{2,3}$, $\Pi_{2,3}$ are obtained in a similar way. The RG equations for the velocities are given by
\bea
	\frac{d v_x}{dl} &=& \frac{g^2}{(2\pi)^d} \left(\frac{16\pi^2(c_x-v_x)}{3c_x(c_x+v_x)^2}+ \frac{2\pi^2(2c_x^2-3c_x v_x+ 7v_x^2)}{3 c_x v_x(c_x+v_x)^3} \delta v- \frac{8\pi^2(c_x^2-3c_x v_x - v_x^2)}{3c_x^2(c_x+v_x)^3} \delta c \right), \\
	\frac{d c_x}{dl} &=& \frac{g^2}{(2\pi)^d} \left(\frac{\pi^2(v_x^2-c_x^2)}{c_x v_x^3}+ \frac{\pi^2(2c_x^2+ v_x^2)}{2c_x v_x^4} \delta v \right), \\
	\frac{d \delta v}{dl} &=& \frac{g^2}{(2\pi)^d} \left( -\frac{4\pi^2(6c_x^2+33c_x v_x+ 31v_x^2)}{15 c_x v_x(c_x+v_x)^3} \delta v- \frac{16\pi^2(c_x^2+3c_x v_x+ v_x^2)}{15 c_x^2 (c_x+v_x)^3} \delta c + \frac{16\pi^2}{3c_x(c_x+v_x)^3} \delta v^2+ \frac{8\pi^2(3c_x+v_x)}{3c_x^2(c_x+v_x)^3} \delta v \delta c \right), \nn \\
	\\
	\frac{d \delta c}{dl} &=& \frac{g^2}{(2\pi)^d}  \left( - \frac{\pi^2(c_x^2 +v_x^2)}{c_x^2 v_x^3} \delta c- \frac{\pi^2}{c_x v_x^2} \delta v+ \frac{\pi^2(2c_x^2-v_x^2)}{2c_x^2 v_x^4} \delta v \delta c \right).
\eea

By solving these equations, we find that there is a line of stable fixed points with $v_x\!=\! c_x \!\equiv\! v$, $\delta v \!=\! \delta c \!=\! 0$. The linearized RG equations around the fixed point $(v,v,0,0)$ read, in matrix form,
\bea
	\frac{d}{dl} \left( \ba{cccc} v_x \\ c_x \\ \delta v \\ \delta c \ea \right)=  \frac{\pi^2 g^2}{v^3}\left( \ba{cccc} -\frac43 & \frac43 & \frac12 & 1 \\ 2 & -2 & \frac32 & 0 \\ 0 & 0 & -\frac73 & -\frac23 \\ 0 & 0 & -1 & -2 \ea\right) \left( \ba{cccc} v_x \\ c_x \\ \delta v \\ \delta c \ea \right).
\eea
The eigenvalues of the RG stability matrix at the fixed point are $\frac{\pi^2 g^2}{v^3}(-\frac{10}{3},-3, -\frac43,0)$. This means the perturbation deviating from the fixed line is irrelevant, while the perturbation along the fixed line is marginal, corresponding to the fact that the value of $v$ is not universal. The fixed point thus has emergent Lorentz symmetry. In the following, we set $v_x\!=\!v_y\!=\!c_x\!=\!c_y\!=\!1$ for simplicity, and the three propagators are equal, namely, $D_i(p)\!=\!D(p)$ and $S_i(p)\!=\!S(p)$ for $i\!=\!1,2,3$. Anomalous dimensions for the boson and fermion fields are given by $\eta_\phi \!=\! \eta_\psi \!=\! \frac{A_{d\!-\!1}}{(2\pi)^d} \frac{\pi}{4} g^2$, where $A_{d\!-\!1}$ is the area of a unit $(d-1)$-sphere. The renormalization of the four-boson vertices is given below, where we work in $4\!-\!\epsilon$ spacetime dimensions and use dimensional regularization:
\bea
	V_b^{(a)}&=& -\frac12 \int_k D(k)^2\Big[ (6u^2+32u'^2+16uu'') ( |\phi_1|^2 |\phi_2|^2+ \text{c.p.})  + (4u'^2+8uu'+8u'u'')[(\phi_1^{\ast2} \phi_2^2+\text{h.c.})+ \text{c.p.}] \nn\\
	&& + (2u^2+8u'^2+20u''^2) (|\phi_1|^4+\text{c.p.}) \Big] \nn\\
	&=& - \frac{A_{d-1}\Lambda^{d-4}}{(2\pi)^d\epsilon} \frac{\pi}{4} \Big[ (6u^2+32u'^2+16uu'') ( |\phi_1|^2 |\phi_2|^2+ \text{c.p.})  + (4u'^2+8uu'+8u'u'') [(\phi_1^{\ast2} \phi_2^2+\text{h.c.})+ \text{c.p.}] \nn\\
	&& + (2u^2+8u'^2+20u''^2) (|\phi_1|^4+\text{c.p.}) \Big], \\
	V_b^{(b)}&=& \frac12 g^4 \int_k \text{Tr}[\sigma^y S(k) \sigma^y S^T(k) \sigma^y S(k) \sigma^y S^T(k)] \Big[2 ( |\phi_1|^2 |\phi_2|^2+ \text{c.p.}) + (|\phi_1|^4+\text{c.p.})\Big] \nn\\
	&=&\frac{A_{d-1}\Lambda^{d-4}}{(2\pi)^d\epsilon} \frac{\pi  g^4}{2 }  [2( |\phi_1|^2 |\phi_2|^2+ \text{c.p.}) +(|\phi_1|^4+\text{c.p.})].
\eea
Note that there is no renormalization of the fermion-boson vertex (\ref{EqS6}).

Absorbing the vertex correction into the running coupling constants, the RG equations read
\bea
	\frac{d g^2}{dl} &=& \epsilon g^2-\frac{3\pi}{2} g^4, \\
	\frac{du}{dl} &=&\epsilon u -\pi g^2 u+ \pi g^4- \frac{\pi}{2} (3u^2+16u'^2+8uu''), \\
	\frac{du'}{dl} &=&\epsilon u'  -\pi g^2 u'- \pi(u'^2+2uu'+2u'u''), \\
	\frac{du''}{dl} &=& \epsilon u'' -\pi g^2 u'' + \frac{\pi}{2} g^4 - \frac{\pi}{2}(u^2+4u'^2+10u''^2),
\eea
where we have rescaled the coupling constants as $g \!\rightarrow\! (\frac{(2\pi)^d}{A_{d-1}})^{1/2} g$, $u_i \!\rightarrow\! \frac{(2\pi)^d}{A_{d-1}} u_i$. After this rescaling, the anomalous dimensions become $\eta_\phi= \eta_\psi=\frac{\pi}{4} g^2$.

Since the flow equation for $g^2$ decouples from the others, one can immediately determine that $g^{*2}\!=\!\frac{2\epsilon}{3\pi}$ is a stable fixed point. Solving the full set of RG equations, we find a fixed point with
\bea\label{EqS25}
(g^{2}_{\text{susy}}, u_{\text{susy}},u^{\prime}_{\text{susy}},u^{\prime\prime}_{\text{susy}})=\left(\frac{2\epsilon}{3\pi},\frac{2\epsilon}{3\pi},0,0\right),
\eea
corresponding to emergent SUSY of the XYZ model type (see Sec.~C below). One obtains the linearized RG equations in the main text by expanding the RG equations near the SUSY fixed point in 2+1 dimensions,
\bea
	\frac{d}{dl} \left( \ba{cccc} \delta g^2 \\ \delta u \\ \delta u' \\ \delta u'' \ea \right) = \left( \ba{cccc} -1 & 0 & 0 & 0 \\ \frac23 & -\frac53 & 0 & -\frac83 \\ 0 & 0 & -1 & 0 \\ \frac23 & -\frac23 & 0 & \frac13 \ea \right) \left( \ba{cccc} \delta g^2 \\ \delta u \\ \delta u' \\ \delta u'' \ea \right),
\eea
where $\delta g^2, \delta u_i$ are deviations from the SUSY fixed point. Note that this fixed point corresponds to a tricritical point and is unstable as explained in the main text. The critical exponent $\eta$ at the SUSY fixed point is $\eta=2\eta_\phi=\frac\epsilon3$. Next, we calculate the correlation length exponent $\nu\equiv \Delta_r^{-1}$ at the SUSY fixed point. The one-loop contribution to the mass term is
\bea
	\Pi= 2u \int_k \frac{1}{k^2+r}=-\frac{A_{d-1}}{(2\pi)^d\epsilon} \pi u r,
\eea
where $r$ is the boson mass. Thus $\Delta_r=2-2\eta_\phi-\frac{A_{d-1}}{(2\pi)^d} \pi u=2-\epsilon$ and
\bea
\nu=\frac12+\frac\epsilon4+O(\epsilon^2).
\eea

From the RG equations, there is only one stable fixed point given by
\bea
(g^{2}_{\text{st}},u_{\text{st}},u^{\prime}_{\text{st}},u^{\prime\prime}_{\text{st}})=\left(\frac{2\epsilon}{3\pi},\frac{1+\sqrt{57}}{21\pi}\epsilon,0,\frac{1+\sqrt{57}}{42\pi}\epsilon\right),
\eea
where there is an emergent SO(6) symmetry in the boson potential. This can be seen from the fact that $u_\text{st}=2u^{\prime\prime}_\text{st}$, and thus at the fixed point
\bea
	V_b= 2u^{\prime\prime}_\text{st} (|\phi_1|^2 |\phi_2|^2 + \text{c.p.}) + u^{\prime\prime}_\text{st} \sum_i |\phi_i|^4= u^{\prime\prime}_\text{st}\left(\sum_{i=1}^3|\phi_i|^2\right)^2.
\eea

\subsection{C. The XYZ supersymmetric field theory}

In this section, we simply introduce the supersymmetric XYZ model~\cite{strassler2003}, and compare it to the $\mathcal{N}=2$ Wess-Zumino model. The dotted convention is used~\cite{wessbook}. A general chiral superfield is given by
\bea	
	\Phi= \phi+ \sqrt{2} \theta \psi + i\theta \sigma^\mu \bar{\theta} \partial_\mu \phi+ \theta^2 f+ \frac{i}{\sqrt{2}} \theta^2 \bar{\theta} \bar{\sigma}^\mu \partial_\mu \psi- \frac{1}{4} \theta^2 \bar{\theta}^2 \partial^\mu \partial_\mu \phi,
\eea
where $\phi$ is a complex boson, $\psi$ is a two-component Dirac fermion, $\theta,\bar{\theta}$ are Grassman-valued two-component spinors and $f$ is a complex scalar auxiliary field. We define $\sigma^0$=$-I$ with $I$ the identity matrix, and $\sigma^i$ are Pauli matrices. Right- (left-)handed invariant tensors are defined as $\sigma^\mu=(\sigma^0,\vec{\sigma})$ [$\bar{\sigma}=(\sigma^0,-\vec{\sigma})]$. Spinor indices are raised (lowered) with the antisymmetric tensor $\varepsilon^{\alpha\beta}$ ($\varepsilon_{\alpha\beta}$). A three-dimensional version of the invariant tensors is obtained by dimension reduction. The kinetic part of the Lagrangian is obtained from the chiral superfields as
\bea\label{Ksuperfield}
	K=\int d^2\bar\theta d^2\theta\, \bar\Phi \Phi= |\partial_\mu \phi|^2- i \bar\psi \bar\sigma^\mu \partial_\mu \psi + ff^*.
\eea
Interaction terms are obtained from the superpotential, which is also expressed in terms of the chiral superfields. The massless $\mathcal{N}=2$ Wess-Zumino model (i.e., at the critical point $r=0$) corresponds to the $\Phi^3$ superpotential,
\bea
	V=-\frac{1}{3} g \int d^2 \theta\, \Phi^3 + \mathrm{h.c.} = -g \phi^2 f+ g \phi \psi^2 + \mathrm{h.c.}
\eea
where $g$ is a complex coupling constant in general. On the other hand, the superpotential for the XYZ model is given by $\Phi_1\Phi_2\Phi_3$,
\bea\label{XYZsuperpotential}
	V= - g \int d^2 \theta\, \Phi_1 \Phi_2 \Phi_3 +\mathrm{h.c.} = g [-( \phi_1 \phi_2 f_3 + \text{c.p.}) + ( \phi_1 \psi_2 \psi_3 + \text{c.p.})] + \mathrm{h.c.}
\eea
Integrating out the auxiliary fields $f,f^*$, the Wess-Zumino model becomes
\bea
	\mathcal{L}_{\text{WZ}}=|\partial_\mu \phi|^2- i \bar\psi \bar\sigma^\mu \partial_\mu \psi+ g (\phi\psi^2 +\phi^* \bar{\psi}^2) - |g|^2 |\phi|^4,
\eea
while the XYZ model becomes
\bea
	\mathcal{L}_{\text{XYZ}} &=& \sum_{i=1}^3 \int d^2\bar\theta d^2\theta\, \bar\Phi_i \Phi_i- g\left( \int d^2 \theta\, \Phi_1 \Phi_2 \Phi_3 +\mathrm{h.c.} \right) \label{xyz_lagrangian} \\
	&=& \sum_{i=1}^3 [|\partial_\mu \phi_i|^2- i \bar\psi_i \bar\sigma^\mu \partial_\mu \psi_i]+ g[( \phi_1 \psi_2 \psi_3+ \mathrm{c.p.})+ \mathrm{h.c.}]- |g|^2 (|\phi_1|^2 |\phi_2|^2+ \mathrm{c.p.}),
\eea
which is precisely the form of our effective Lagrangian at the fixed point (\ref{EqS25}), corresponding to the NPDW tricritical point, if one sets $g \rightarrow i g$.

\subsection{D. Mirror symmetry: XYZ model and $\mathcal{N}=2$ SQED with $N_f=1$}

In this section we simply state the infrared duality in 2+1 dimensions between the XYZ model and $\mathcal{N}=2$ SQED with two matter chiral multiplets $(Q,\tilde{Q})$ (referred to in the SUSY literature as a single flavor of matter fields, $N_f=1$). This duality is known as $\mathcal{N}=2$ mirror symmetry. For a more detailed discussion of mirror symmetry that is nonetheless accessible to non-specialists, we direct the reader to Ref.~\cite{strassler2003,hook2013,hook2014}.

The field content of $\mathcal{N}=2$ SQED with $N_f=1$ is a vector multiplet $V$ and two chiral multiplets $Q$ and $\tilde{Q}$. The vector multiplet plays the role of the gauge field in the usual particle-vortex duality, and contains a bosonic gauge field $A_\mu$, a real dynamical scalar field $\sigma$, a real auxiliary scalar field $D$, and a two-component fermionic gaugino $\lambda$. The chiral multiplets are matter fields and contain each a complex scalar, a two-component Dirac fermion, and a complex scalar auxiliary field (see Sec.~C above). In the superfield formalism, the Lagrangian is given by
\bea\label{LSQED}
\mathcal{L}_\text{SQED}=\frac{1}{2g^2}\int d^2\theta\,W_\alpha W^\alpha+\text{h.c.}+\int d^2\bar{\theta}d^2\theta\left(Q^\dag e^{2V}Q+\tilde{Q}^\dag e^{-2V}\tilde{Q}\right),
\eea
where $W_\alpha$ is the gaugino multiplet that can be generated from the vector multiplet by taking derivatives. The first two terms are the SUSY analog of the Maxwell term, and read in component form
\bea
\frac{1}{2g^2}\int d^2\theta\,W_\alpha W^\alpha+\text{h.c.}=\frac{1}{g^2}\left(-\frac{1}{4}F_{\mu\nu}^2+\frac{1}{2}(\partial_\mu\sigma)^2-i\bar{\lambda}\slashed{\partial}\lambda+\frac{1}{2}D^2\right),
\eea
where $g$ is the gauge coupling and $F_{\mu\nu}=\partial_\mu A_\nu-\partial_\nu A_\mu$ is the field strength. The last term in (\ref{LSQED}) corresponds to the SUSY analog of minimal coupling between the matter fields $Q,\tilde{Q}$ and the vector field $V$. The theory (\ref{LSQED}) flows in the infrared to a strongly coupled fixed point, which is the same fixed point as that of the dual Lagrangian \cite{aharony1997}
\bea
\mathcal{L}_\text{dual}=\int d^2\bar{\theta}d^2\theta\left(\bar{V}_+ V_+ +\bar{V}_- V_-+ \bar{M} M\right)-g\left(\int d^2\theta\,MV_+V_-+\mathrm{h.c.}\right),
\eea
where $M=Q\tilde{Q}$ and $V_\pm$ are three chiral superfields ($V_\pm$ originate from $V$). Comparing with Eq.~(\ref{xyz_lagrangian}), the dual Lagrangian is identified with the XYZ model. Thus the XYZ model and $\mathcal{N}=2$ SQED with $N_f=1$ have identical low-energy/long-wavelength properties.

\subsection{E. Exact evaluation of the zero-temperature optical conductivity}

In this section, we follow the derivation in Ref. \cite{wwk2015} closely to evaluate the transport properties at the tricritical theory with the emergent SUSY of XYZ/SQED type. According to the Kubo formula, the longitudinal optical conductivity is given by the current-current correlation function,
\bea
	\sigma(\omega)= \frac{e^2}{\hbar}\frac{1}{i\omega}\langle J_x(\omega) J_x(-\omega) \rangle,
\eea
where $\omega$ is frequency, $\frac{e^2}{\hbar}$ is the conductivity quantum and $J_x(\omega)$ is the current operator at zero spatial momentum. At the SUSY fixed point, the system respects conformal symmetry, which highly constrains the two-point function of the U(1) current even in $2\!+\!1$ dimensions~\cite{osborn1994}. We obtain
\bea
	\langle J_\mu(x) J_\nu(0) \rangle &=& C_{J} \frac{I_{\mu\nu}(x)}{x^{2(d-1)}},
\eea
where $\mu,\nu$ are space-time indices, $d$ is the spacetime dimension, $C_J$ is a positive coefficient, and $I_{\mu\nu}(x)$ is defined as
\bea
	I_{\mu\nu}(x)&=& \delta_{\mu\nu}-2 \frac{x_\mu x_\nu}{x^2}. \label{JJ}
\eea

In addition to conformal symmetry, the tricritical theory has emergent $\mathcal{N}=2$ SUSY. It has an additional $R$-symmetry~\cite{wessbook}. As a result the two-point function of the $R$-current is related to the two-point function of the U(1) current by~\cite{nishioka2013, wwk2015} $C_J\!=\!\frac{5\tau_{RR}}{2\pi^2}$, where $\tau_{RR}$ is the coefficient of the two-point function of the $R$-current.

Fourier transforming Eq.~(\ref{JJ}) and using the $R$-symmetry, the zero-temperature optical conductivity is given by~$\sigma_0(\omega)=\frac54 \tau_{RR}\frac{e^2}{\hbar}$~\cite{wwk2015}. Note that it is independent of frequency owing to conformal symmetry~\cite{damle1997}. The coefficient $\tau_{RR}$ is a function of the $R$-charge of the chiral superfield~\cite{nishioka2013},
\bea
	\tau_{RR}(\mathcal{R}_I)= \sum_I \frac{2}{\pi^2} \int_0^\infty dx \left[ (1-\mathcal{R}_I)\left( \frac{1}{x^2} -\frac{\cosh(2x(1-\mathcal{R}_I))}{\sinh^2 x} \right)+ \frac{(\sinh 2x -2x) \sinh(2x(1-\mathcal{R}_I))}{2\sinh^4 x}\right],
\eea
where $\mathcal{R}_I \equiv \mathcal{R}(\Phi_I)$ is the $R$-charge of the chiral superfield $\Phi_I$. In our case, owing to the C$_3$ rotational symmetry relating the three chiral superfields, we have
\bea
\mathcal{R}_1=\mathcal{R}_2=\mathcal{R}_3, \\
\mathcal{R}_1+\mathcal{R}_2+\mathcal{R}_3=2,
\eea
hence $\mathcal{R}_1\!=\!\mathcal{R}_2\!=\!\mathcal{R}_3\!=\!\frac23$. As a result, $\tau_{RR}$ in the XYZ model is simply three times that of the Wess-Zumino model, which is evaluated analytically in Ref.~\cite{wwk2015}. Thus we have $\tau_{RR}=3\times \frac{4}{243}(16-\frac{9\sqrt{3}}{\pi}) \approx 0.545$ and
\bea
\sigma_0(\omega)=\frac{15}{243}\left(16-\frac{9\sqrt{3}}{\pi}\right) \frac{e^2}{\hbar} \approx 0.681 \frac{e^2}{\hbar}.
\eea

\end{widetext}

\end{document}